\documentclass{article}
\usepackage[utf8]{inputenc}
\usepackage{amsmath, amssymb}

\usepackage{mathtools}                  %
\mathtoolsset{showonlyrefs=false}        %
\usepackage[normalem]{ulem}
\usepackage{mathabx}
\DeclareMathOperator*{\argmin}{arg\,min}
\DeclareMathOperator*{\itm}{ITM}
\usepackage[ruled,vlined]{algorithm2e}
\usepackage{csquotes}

\usepackage{ulem}

\usepackage{placeins}  %

\usepackage{xcolor}
\usepackage[english]{babel}
\usepackage[textsize=footnotesize]{todonotes}

\DeclareMathOperator*{\amax}{\mathnormal{\alpha}-max}

\usepackage{tikz}
\usepackage{tikz-qtree}
\usetikzlibrary{calc}
\usetikzlibrary{decorations}
\tikzset{
    cheating dash/.code args={on #1 off #2}{
        \csname tikz@addoption\endcsname{%
            \pgfgetpath\currentpath%
            \pgfprocessround{\currentpath}{\currentpath}%
            \csname pgf@decorate@parsesoftpath\endcsname{\currentpath}{\currentpath}%
            \pgfmathparse{\csname pgf@decorate@totalpathlength\endcsname-#1}\let\rest=\pgfmathresult%
            \pgfmathparse{#1+#2}\let\onoff=\pgfmathresult%
            \pgfmathparse{max(floor(\rest/\onoff), 1)}\let\nfullonoff=\pgfmathresult%
            \pgfmathparse{max((\rest-\onoff*\nfullonoff)/\nfullonoff+#2, #2)}\let\offexpand=\pgfmathresult%
            \pgfsetdash{{#1}{\offexpand}}{0pt}}%
    }
}

\usepackage{hyperref}
\hypersetup{
    colorlinks,
    linkcolor={red!50!black},
    citecolor={blue!50!black},
    urlcolor={blue!80!black}
}
\usepackage{cleveref}

\usepackage{zref-savepos}
\usepackage{pict2e}
  \usepackage[
    style=alphabetic,
  ]{biblatex}

\newcounter{NoTableEntry}
\renewcommand*{\theNoTableEntry}{NTE-\the\value{NoTableEntry}}

\usepackage{mathtools}
\newcommand{\mbb}[1]{\mathbb{#1}}
\newcommand{\mcal}[1]{\mathcal{#1}}
\DeclareMathOperator*{\esssup}{ess\,sup}

\DeclarePairedDelimiter{\pars}{\ensuremath{(}}{\ensuremath{)}}
\DeclarePairedDelimiter{\bracs}{\ensuremath{[}}{\ensuremath{]}}
\DeclarePairedDelimiter{\braces}{\ensuremath{\{}}{\ensuremath{\}}}

\DeclarePairedDelimiter{\norm}{\|}{\|}
\DeclarePairedDelimiter{\abs}{\lvert}{\rvert}
\renewcommand{\vec}[1]{\mathbf{#1}}
\newcommand*{\dx}[1]{\ensuremath{\,\mathrm{d}{#1}}}

\makeatletter
\newcommand*{\notableentry}{%
  \kern-\tabcolsep
  \stepcounter{NoTableEntry}%
  \vadjust pre{\zsavepos{\theNoTableEntry t}}%
  \vadjust{\zsavepos{\theNoTableEntry b}}%
  \zsavepos{\theNoTableEntry l}%
  \raisebox{%
    \dimexpr\zposy{\theNoTableEntry b}sp
    -\zposy{\theNoTableEntry l}sp\relax
  }[0pt][0pt]{%
    \color{black}%
    \setlength{\unitlength}{1pt}%
    \edef\w{%
      \strip@pt\dimexpr\zposx{\theNoTableEntry r}sp%
      -\zposx{\theNoTableEntry l}sp\relax
    }%
    \edef\h{%
      \strip@pt\dimexpr\zposy{\theNoTableEntry t}sp%
      -\zposy{\theNoTableEntry b}sp\relax
    }%
    \ifdim\w pt=0pt %
    \else
      \begin{picture}(0,0)%
        \edef\x{%
          \noexpand\put(0,0){\noexpand\line(\w,\h){\w}}%
          \noexpand\put(0,\h){\noexpand\line(\w,-\h){\w}}%
        }\x
      \end{picture}%
    \fi
  }%
  \hspace{0pt plus 1filll}%
  \zsavepos{\theNoTableEntry r}%
  \kern-\tabcolsep
}

\title{Pricing high-dimensional Bermudan options with hierarchical tensor formats}

\renewcommand\@date{{%
  \vspace{-\baselineskip}%
  \large\centering
  \begin{tabular}{@{}c@{}}
    Christian Bayer\textsuperscript{1} \\
    \normalsize \texttt{christian.bayer@wias-berlin.de} 
  \end{tabular}%
  \quad 
  \begin{tabular}{@{}c@{}}
    Martin Eigel\textsuperscript{1} \\
    \normalsize \texttt{martin.eigel@wias-berlin.de} 
  \end{tabular}
  \quad 
  \begin{tabular}{@{}c@{}}
    Leon Sallandt\textsuperscript{2} \\
    \normalsize \texttt{sallandt@math.tu-berlin.de} 
  \end{tabular}
  \quad 
  \begin{tabular}{@{}c@{}}
    Philipp Trunschke\textsuperscript{2} \\
    \normalsize \texttt{ptrunschke@mail.tu-berlin.de} 
  \end{tabular}

  \bigskip

  \textsuperscript{1}Weierstrass Institute for Applied Analysis and Stochastics\par
  \textsuperscript{2}Technische Universit\"at Berlin

  \bigskip

  \today
}}

\begin{document}

\maketitle

\begin{abstract}
An efficient compression technique based on hierarchical tensors for popular option pricing methods is presented.
It is shown that the ``curse of dimensionality'' can be alleviated for the computation of Bermudan option prices with the Monte Carlo least-squares approach as well as the dual martingale method, both using high-dimensional tensorized polynomial expansions.
This discretization allows for a simple and computationally cheap evaluation of conditional expectations.
Complexity estimates are provided as well as a description of the optimization procedures in the tensor train format.
Numerical experiments illustrate the favourable accuracy of the proposed methods.
The dynamical programming method yields results comparable to recent Neural Network based methods.
\end{abstract}

\section{Introduction}
\label{sec:introduction}

Pricing of American or Bermudan type options, i.e., options with an early exercise feature, is one of the most classical, but also most difficult problems of computational finance, producing a vast amount of literature. Some examples of popular classes of methods include PDE methods (see, for instance, \cite{achdou2005computational}), tree and stochastic mesh methods (see, for instance, \cite{glasserman2013monte}), and policy iteration (see, e.g., \cite{belomestny2018advanced}). In this paper we consider two other very popular methodologies, namely least squares Monte Carlo methods based on the dynamic programming principles pioneered by \cite{longstaff2001valuing} and dual martingale methods introduced by \cite{rogers2002MCOptions}, both of which were, of course, widely adapted and considerably improved since then. We refer to \cite{ludkovski2020mlosp} for a recent overview together with an open-source implementation.

Both least squares Monte Carlo methods and duality methods require efficient and accurate approximation of functions from a potentially large class. Indeed, the key step of the least squares Monte Carlo method involves the computation of a \emph{continuation value}, i.e., of the conditional expectation $\mathbb{E}_t[v(t+\Delta t, X_{t+\Delta t})]$ of a future \emph{value function} at time $t$.\footnote{Actual algorithms may rather involve actual future payoffs such as in \cite{longstaff2001valuing}. Note that we ignore discounting at this time.} (For sake of presentation, let us assume that we are using an asset price model based on a Markov process $X$, which contains the asset prices $S$, but possibly also further components, such as stochastic volatilities or interest rates.) This conditional expectation is then approximated within a finite dimensional space spanned by \emph{basis functions} -- often chosen to be polynomials. When the dimension $d$ of the underlying process $X$ is high, we encounter a curse of dimensionality, i.e., we expect that the number of basis functions needed to achieve a certain accuracy increases exponentially in the dimension $d$. This is especially true when the basis functions are chosen by ``tensorization'' of one-dimensional basis functions. E.g., the dimension of the space of polynomials of (total) degree $p$ in $d$ variables is $\binom{d+p}{d}$. Such a polynomial basis become inefficient when $d \gg 1$, a realistic scenario for options on baskets or indices. For instance, options on SPY (with $100$ assets) are American, implying that $d \ge 100$, depending on the choice of the model -- in the sense that continuation values also depend on volatilities not just the asset prices in stochastic volatility models, for example. Hence, other classes of basis functions are needed.

Duality methods are typically based on parameterizations of families of candidate martingales. In the Markovian case, we may restrict ourselves to martingales representable as stochastic integrals of functions $\phi(t,X_t)$ against the driving Brownian motion, and we again see a potential curse of dimensionality in terms of the dimension of $X$.

When the underlying model is \emph{not Markovian} -- as, e.g., common for \emph{rough volatility models}, see, e.g., \cite{bayer2016pricing} -- the involved dimensions can increase drastically, as then both continuation values and candidate martingales theoretically depend on the entire trajectory of the process $X$ until time $t$. There are only very few rigorously analyzed methods for such non-Markovian problems. We specifically refer to \cite{lelong2018dual,lelong2019pricing}, both of which are based on Wiener chaos expansions of the value process and the candidate martingale, respectively. In this framework, conditional expectations can be computed explicitly, but the curse of dimension enters via the chaos decomposition itself, see Section~\ref{sec:dual} for details.

In either case, we are faced with ``natural'' $d$-dimensional bases which quickly increase in size as $d$ increases. While the curse of dimension is often a real, inescapable fact of complexity theory (in the sense of a worst case dependence over sufficiently general classes of approximation problems), real life problems often exhibit structural properties which lead to a notion of ``effective dimension'' of a problem which may increase much slower than the actual dimension $d$ -- see, for instance, \cite{wang2005high} for a similar phenomenon in finance. This insight has lead to efficient approximation strategies for high-dimensional functions of low effective dimension of some sort in numerical analysis. In this paper, we propose to use hierarchical tensor formats, more precisely \emph{tensor trains}, to provide efficient approximations of nominally high-dimensional functions, provided that they allow for accurate \emph{low-rank approximations}.

Hierarchical tensors (HT)~\cite{bachmayr2016tensor,hackbusch2014tensor} rely on the classical concept of~\textit{separation of variables} by means of a generalization of the singular value decomposition (SVD) to higher-order tensors, preserving many of its well-known properties.
The hierarchical SVD (HSVD) yields a notion of multilinear rank and provides an approach to obtain a quasi-optimal low-rank approximation by rank truncation.
For fixed multilinear ranks, the representation and operation complexities of these formats scale only linearly in the order of the tensor.
Central to the HSVD is a tree-based representation of a recursive decomposition of the tensor space into nested subspaces.
For the described algorithms, we use the common tensor train (TT) format~\cite{Oseledets2009,Oseledets-2011,oseledets2013constructive}, which is a ``linearization'' of the HT representation with general binary trees.
Similar to matrices, the set of hierarchical tensors of fixed multilinear rank is not convex but forms a smooth manifold.
Hence, appropriate optimization techniques such as alternating and Riemannian schemes are available.

Tensor trains are a new technique in computational finance. In fact, we are only aware of one other paper in the field using these tensor representations, namely \cite{glau2020low}. In that paper, the authors consider parametric option pricing problems. That is, they are given a model with parameters $\zeta$ and options with parameters $\eta$. The price of these options in the model is then a function $P(\theta)$, $\theta \coloneqq (\zeta, \eta)$, of the model and option parameters, and we can expect $P$ to be regular. Some tasks in financial engineering require rapid option pricing, e.g., for calibrating model parameters to market prices. Following \cite{gass2018chebyshev}, \cite{glau2020low} propose to approximate $\theta \mapsto P(\theta)$ by Chebyshev interpolation. If $\theta$ is high-dimensional, such a interpolation may already involve a very large number of Chebyshev polynomials, and they then proceed to ``compress'' the representation using tensor trains.

No discussion of computational methods for high-dimensional problems can today ignore the trend of using machine learning techniques, in particular deep neural networks, to often great success. In the context of American or Bermudan options, we mention the recent paper by \cite{becker2019deep}, who are able to accurately price high-dimensional Bermudan options in dimensions up to $500$ using deep learning techniques based on parameterization of \emph{randomized} stopping times, see also \cite{bayer2020pricing}. A natural question then is if the successes of deep learning for solving high dimensional problems (\emph{``overcoming'' the curse of dimension}) can also be achieved by other, more traditional methods of numerical analysis. 

\subsection*{Main contributions}
\label{sec:contributions}

Our intention is to advocate the use of hierarchical tensor formats for high-dimensional problems in computational finance.
For this, we provide an overview of the main ideas of these formats and illustrate the application of tensor trains with two popular methods using tensorized polynomial spaces for the discretization.
The considered problem sizes would be infeasible without some efficient model order reduction technique.
We demonstrate in particular that the achieved accuracy is comparable to recent Neural Network approaches.

Tensor networks have already been used to alleviate the curse of dimensionality in physics~\cite{vidal2003mps}, parametric PDEs~\cite{bachmayr2016tensor,eigel2017adaptive,eigel2019variational,eigel2020lognormal} as well as other control problems~\cite{dolgov2019tensor, oster2019approximating, fackeldey2020approximative}.
They may significantly reduce the computational complexity~\cite{hackbusch2012book} and are able to represent sparse functions with a constant overhead~\cite{bachmayr2017sparseVsLowrank}.
In this paper we demonstrate the usefulness of tensor networks in computational finance on two examples with discretizations in polynomial tensor product spaces in $d$ dimensions with degree $p$ of the form
\begin{equation}
    X = \sum_{\alpha \in [p]^d} X_\alpha P_\alpha
\end{equation}
with coefficient tensor $X\in\mathbb R^{p^d}$.
The first example showcases the application of the alternating least squares algorithm~\cite{Holtz2012a} for the best approximation problem in the primal method of Longstaff and Schwartz~\cite{longstaff2001valuing} where the discounted value is given by
\begin{equation}
    v(x) = \sum_{\alpha\in\Lambda} V_\alpha \prod_{k=1}^{d'} B_{\alpha_k}(x_k).
\end{equation}

In the second example we present the application of a Riemannian optimization algorithm~\cite{Kressner2014} to solve the convex minimization problem in the dual method of Lelong~\cite{lelong2018dual}.
For both examples we examine the reduction of the space and time complexity.
In the numerical experiments we compare the originally published and the new methods on standard problems. 
The reduced complexity allows to apply the Longstaff-Schwartz algorithm to problems with up to $1000$ assets.
Problems of this size have only been reported recently with state-of-the-art machine learning methods~\cite{becker2019deep}.
Moreover, in comparison to the Neural Network approach, our method requires significantly fewer samples.
Even though the application of the tensor compression to the dual method turned out to be quite involved (in terms of the tensor optimization), the resulting algorithm produces comparable or better results while considerably reducing the dimensionality of the underlying equation.
This renders this approach tractable for more assets and higher accuracy computations.

We conclude that tensor networks can be very beneficial technique for high-dimensional problems in financial mathematics.
They rival the performance of Neural Networks, show similar approximation and complexity properties, and exhibit richer mathematical structures that can be exploited (such as in the Riemannian optimization described in Section~\ref{sec:differentiable}).

\section{Bermudan option pricing}
\label{sec:option pricing}

In what follows we introduce our frameworks and notations for the Bermudan option pricing problem. Furthermore, we recall the celebrated Longstaff-Schwartz algorithm as well as Lelong's version of Rogers' duality approach based on a Wiener chaos expansion.

We fix some finite time horizon $T>0$ and a filtered probability space $\pars{\Omega, \mcal{F}, \pars{\mcal{F}_t}_{0\le t\le T}, \mbb{P}}$, where $\pars{\mcal{F}_t}_{0\le t\le T}$ is supposed to be the natural augmented filtration of a $d$-dimensional Brownian motion $B$ -- the natural setting for the Wiener chaos expansion lying at the core of our duality algorithm.
On this space, we consider an adapted Markov process $\pars{S_t}_{0\le t\le T}$ with values in $\mbb{R}^{d'}$ modeling a $d'$-dimensional underlying asset.
The number of assets $d'$ can be smaller than the dimension $d$ of the Brownian motion to encompass the case of stochastic volatility models or stochastic interest rate.
To simplify notation, we consider the case that $S$ generates the filtration and $d' = d$.

We assume that %
$\mbb{P}$ is an associated risk neutral measure.
We consider an adapted payoff process $\widetilde{Z}$ and introduce its discounted value process
\[
\pars*{Z_t = \exp\pars{-\int_0^t r\pars{s} \dx{s}} \widetilde{Z}_t}_{0\le t\le T}.
\]
We assume that the paths of $Z$ are right continuous and that $\sup_{t\in\bracs{0,T}} \abs{Z_t} \in L^2(\Omega, \mathcal{F}_T, \mathbb{P})$.
The process $\widetilde{Z}$ can obviously take the simple form $\pars{\varphi\pars{S_t}}_{t\le T}$ for some function $\varphi$,
but it can also depend on the whole path of the underlying asset $S$ up to the current time. %
We consider the Bermudan option paying $\widetilde{Z}_{t_k}$ to its holder if exercised at time $0 = t_1 < \dots < t_N = T$.
Standard arbitrage pricing theory defines the discounted time-$t$ value of the Bermudan option to be
\begin{equation}\label{eq:valuefun}
    U_{t_n} = \esssup_{\tau\in\mcal{T}_{t_n}} \mbb{E}\bracs{Z_\tau \vert \mcal{F}_{\tau}}
\end{equation}
where $\mcal{T}_t$ denotes the discrete set of $\mcal{F}$-stopping times with values in $\bracs{t,T}$.

We now recall two of the many algorithms for pricing Bermudan options available in the literature, beginning with the classical Longstaff-Schwartz algorithm. These algorithms will be used to test the efficiency gains achievable by hierarchical tensor formats in the context of option pricing.

\subsection{Primal (Longstaff-Schwartz)}
\label{sec:primal}

In the Longstaff-Schwartz algorithm \cite{longstaff2001valuing}, the dynamic programming principle corresponding to the discounted time-$t$ value of the Bermudan option \eqref{eq:valuefun}, is used. It reads
\begin{equation}\label{eq:bellman}
    U_{t_n} = \max \{ Z_{t_n}, \mathbb E [U_{t_{n+1}} | \mathcal F_{t_n} ] \}
\end{equation}
with final condition $U_{t_N} = Z_{t_N}$.
If $\mathbb E[U_{t_n+1} | \mathcal F_{t_n}]$ is known, an optimal stopping-time policy can be synthesized explicitly by stopping if and only if $Z_{t_n} \geq \mathbb E[U_{t_{n+1}} | \mathcal F_{t_n}]$.
Thus, the problem of finding the optimal stopping time and also the valuation of the option can be reduced to finding $\mathbb E[U_{t_{n+1}} | \mathcal F_{t_n}]$, which is exactly what the Longstaff-Schwartz algorithm approximates.
As this algorithm is pretty standard, we do not give a detailed explanation and instead simply state the algorithm.
Note that we abbreviate the notation by dropping the $t$ in the discretization, i.e. $S_{t_n} = S_n$.
We define the $\itm$ (``in the money'') operator which is mapping a set of assets to the subset where the current payoff is positive.
\begin{algorithm}[H]\label{alg:LS}
\SetAlgoLined
\caption{Longstaff-Schwartz}
\SetKwInOut{Input}{input}\SetKwInOut{Output}{output}
\SetKwInOut{Output}{output}\SetKwInOut{Output}{output}
\Input{Number of samples $M$, exercise dates $0 = t_1 < \dots < t_N = T$, initial value $s_0$.}
\Output{Conditional expectations $ v_n(x) = \mathbb E[U_{n+1} | S_n=x]$, $n \leq N$.}
Set $S_0^m = s_0$ and compute trajectories: $S_n^m$ for $m=1, \dots M$, $n = 1, \dots N$.

Set 
\begin{equation}
Y^m = Z_n^m
\end{equation}

\For{$k = n-1$ to $1$}{
Find $\itm$ paths $S_{\tilde m}$ for $m \in \itm \subset \{1, \dots, M\}$.
Set
\begin{equation}\label{eq:longstaff_regression}
    v_n(\cdot) \approx \argmin_{v \in \mathcal M} \frac 1 {|\text{ITM}|} \sum_{\tilde m \in \text{ITM}} | v(S_n^{\tilde m}) - Y^{\tilde m} |^2.
\end{equation}
    \For{$m = 1$ to $M$}{
        \uIf{$m \in \itm$ and $Z_n^m > v_k(S_n^m)$}{
            $Y^n = Z_n^m$.
        }
    }
}
    Set $v_0(s_0) = \sum_{m = 1}^M Y^m$.
\end{algorithm}
Note that in this formulation of the algorithm, the set $\mathcal M$ in \eqref{eq:longstaff_regression} is traditionally a linear space of polynomials. 
Adding the payoff function to the ansatz space is a common trick to improve the result, see e.g.~\cite{glasserman2013monte}.
In this work we use the set of tensor trains, which we explain in Section~\ref{sec:dual minimization}.

The key computational challenge is the approximation of the conditional expectation
\begin{equation}
    v(S_n) = \mathbb{E}[U_{n+1}|S_n] = \sum_{\alpha\in\mathbb{N}^{d'}} v_\alpha B_\alpha(S_n)
\end{equation}
for some $L^2(\mathbb{R}^{d'}, \mathcal{B}(\mathbb{R}^{d'}), S_*\mathbb{P})$-orthogonal basis $\{B_k\}_{k\in\mathbb{N}}$, where we tacitly assume the payoff having finite second moments.
Since this is an $L^2$-orthogonal projection we can choose a finite set of multi-indices $\Lambda\subset\mathbb{N}^{d'}$ and approximate $\mathbb{E}[Y|S_n]$ by minimizing
\begin{equation}
    \left\| Y - \sum_{\alpha\in\Lambda} v_\alpha B_\alpha(S_n) \right\|^2
    \approx \frac{1}{m}\sum_{i=1}^m \left(Y^m - \sum_{\alpha\in\Lambda} v_\alpha B_\alpha(S_n^m)\right)^2
    . \label{eq:cond_exp_approx}
\end{equation}
We use the index set $\Lambda = [p]^{d'}$ and mitigate the ``curse of dimensionality'' by representing $v$ in the tensor train format as defined in Section~\ref{sec:low-rank tensors}.

\subsection{Chaos-martingale minimization}
\label{sec:dual}

Rogers \cite{rogers2002MCOptions} reformulates the problem of computing $U_{0}$ as the following dual optimization problem
\begin{equation*}
    U_0 = \inf_{M\in H^2_0} \mbb{E}\bracs*{\max_{n=1,\ldots,N} \pars{Z_{t_n} - M_{t_n}}}
\end{equation*}
where $H^2_0$ denotes the set of square integrable martingales vanishing at zero.
This approach requires us to optimize over the space of all (square integrable) martingales. As any martingale $M$ can be expressed as conditional expectations $t \mapsto \mathbb{E}[X|\mathcal{F}_t]$ for some square integrable random variable $X$, we may equivalently solve
\begin{equation} \label{eq:dual_cont}
    U_0 = \inf_{X\in L^2_0\pars{\Omega, \mcal{F}_T,\mbb{P}}} \mbb{E}\bracs*{\max_{n=1,\ldots,N} \pars{Z_{t_n} - \mbb{E}\bracs{X \vert \mcal{F}_{t_n}}}},
\end{equation}
where $L^2_0\pars{\Omega, \mcal{F}_T,\mbb{P}}$ is the set of square integrable $\mcal{F}_T$-random variables with zero mean.
This allows us to minimize over a (seemingly) simpler space -- namely the space of square integrable random variables rather than the space of martingales -- at the cost of expensive calculations of conditional expectations.

The ingenious idea of Lelong \cite{lelong2018dual} was to use a specific parameterization of the space of square integrable random variables in which conditional expectations w.r.t.~the filtration $(\mathcal{F}_t)$ can be computed explicitly at virtually no cost.
Indeed, a finite-dimensional approximation of $X\in L^2_0\pars{\Omega, \mcal{F}_T,\mbb{P}}$ with the above property is given by the truncated Wiener chaos expansion
\begin{equation}
    \label{eq:truncated Wiener}
    \widetilde{X} = \sum_{\alpha\in\Lambda} \widetilde{X}_\alpha H_\alpha\pars{G_1,\ldots,G_N},
\end{equation}
where $\Lambda \subseteq \mathbb{N}^{N\times d'}$ is a predefined set of multi-indices, $H_\alpha$ is the tensorized Hermite polynomial with multi-index $\alpha$ and $G_1, \ldots, G_N$ are $d'$-dimensional Gaussian increments.
The tensorized Hermite polynomials are defined by
\begin{equation}
    H_{\alpha}\pars{G_1,\ldots,G_N} := \prod_{n=1}^N\prod_{k=1}^{d'} h_{\alpha_{nk}}\pars{G_{n,k}}
\end{equation}
where $h_{\alpha_{nk}}$ are the univariate Hermite polynomials with index $\alpha_{nk}$.
Defining the subset $\Lambda^n := \{\alpha\in\Lambda : \forall k>n, \alpha_k=0\}$ it is easy to see that
\begin{equation}
    \mbb{E}\bracs{\widetilde{X} \vert \mcal{F}_{t_n}} = \sum_{\alpha\in\Lambda^n} \widetilde{X}_\alpha H_{\alpha}\pars{G_1, \ldots, G_N} .
\end{equation}
This means that the linear expectation operator $\mbb{E}\bracs{\,\bullet\, \vert \mcal{F}_{t_n}}$ can be represented with the coefficient tensor simply by dropping trailing terms of the chaos expansion.
The expectation in~\eqref{eq:dual_cont} can thus be estimated by the sample average
\begin{equation}
    U_0 = \inf_{\substack{\widetilde{X}_0 = 0 \\ \widetilde{X}_\alpha \in \mbb{R}}} \frac{1}{m}\sum_{i=1}^m \bracs*{\max_{n=1,\ldots, N} \pars*{Z^{\pars{i}}_{t_n} - \sum_{\alpha\in\Lambda^n} \widetilde{X}_\alpha H_{\alpha}\pars{G_1^{\pars{i}}, \ldots, G_N^{\pars{i}}}}} \label{eq:dual_disc},
\end{equation}
where $(Z^{(i)}, G^{(i)})_{1\le i\le m}$ are i.i.d.\ samples from the distribution of $(Z, G)$.
It is shown in~\cite{lelong2018dual} that this is an infimum of a convex, continuous and piece-wise linear cost function over a convex domain and can be calculated easily by a gradient descent descent method with an Armijo line search.

The choice of the multi-index set $\Lambda$ plays an important role in the preformance and applicability of this algorithm.
In~\cite{lelong2018dual} $\Lambda$ is chosen such that the polynomial degree $\sum_{n=1}^N\sum_{k=1}^{d'} \alpha_{nk}$ is bounded by $p$.
This bounds the number of entries of $\widetilde{X}$ that have to be stored by $\binom{Nd'+p}{Nd'} \in \mathcal{O}\left(\frac{(Nd'+p)^p}{p!}\right)$. %
For fixed $p$ this can scale unfavourably when the number of exercise dates $N$ or the dimension of the Brownian motion (i.e.\ the number of assets) $d'$ increases.
We propose to choose $\Lambda = \Lambda_p^N$ such that $\sum_{k=1}^{d'} \alpha_{nk}\le p$ for $\alpha_n\in\Lambda_p$.
and to use the tensor train format to alleviate the ensuing ``curse of dimensionality''.
We introduce the relevant notions and central concepts in the following section.

\section{Low-rank tensor representations}
\label{sec:low-rank tensors}

We are concerned with an efficient representation of expansions of the form $\sum_{\alpha\in\Lambda}U_\alpha\prod_{j=1}^d P_{\alpha_j}$ in tensorized polynomials $P_\alpha$ determined by some finite set $\Lambda\subset \mathcal F := \{\alpha\in\mathbb R^\mathbb N\;:\; |\mathrm{supp} \,\alpha|<\infty \}$ of finitely supported multi-indices.
This representation is used for the considered algorithms with tensorized expansions given by~\eqref{eq:truncated Wiener} and~\eqref{eq:approximate value}.
The set $\Lambda$ typically is given as a tensor set $\Lambda = \bigtimes_{j=1}^d \mathcal I_n:=[n]^d$ or as anisotropical set $\Lambda = \bigtimes_{j=1}^d \mathcal I_{p_j}$, where in our setting $p_j$ denotes the maximal polynomial degree in dimension $j=1,\ldots,d$.
Apparently, $\# \Lambda$ is in $\mathcal O(p^d)$ with $p:=\max\{p_j\;:\; j=1,\ldots,d\}$.
To cope with this exponential complexity, a potentially very efficient approach is the use of low-rank tensor representations as e.g. presented in~\cite{hackbusch2014tensor,nouy2017low}.
Since these modern model reduction techniques are not widely known in the finance community yet, we provide a brief review in order to elucidates some of the central principles.
In the presentation, we follow~\cite{rauhut2017low,bachmayr2016tensor}.

\subsection{Tensor product spaces and subspace approximation}
\label{sec:subspace approximation}

We consider finite dimensional linear spaces $U_i=\mathbb R^{p_i}$ and define the tensor product space
\begin{equation}
    \mathcal H_d := \bigotimes_{j=1}^d U_j.
\end{equation}
Fixing the canonical basis for all $U_j$, any tensor $\mathbf u\in\mathcal H_n$ can be represented by
\begin{equation}
    \label{eq:u expansion}
    \mathbf u = \sum_{\nu_0=1}^{p_1}\cdots\sum_{\nu_n=0}^{p_n} \mathbf U(\nu_1,\ldots,\nu_n)\mathbf e_{\nu_1}^1\otimes\cdots\otimes \mathbf e_{\nu_n}^n,\quad \mathbf U\in\mathbb R^{p_1}\otimes\cdots\otimes\mathbb R^{p_n}.
\end{equation}
Hence, given this basis, any multi-index $\nu\in\mathcal F$ can be identified with a component in the (coefficient) tensor $\mathbf U$, i.e.
\begin{equation}
    \nu=(\nu_1,\ldots,\nu_n) \mapsto \mathbf U(\nu_1,\ldots,\nu_n)\in\mathbb R.
\end{equation}
The goal is to obtain a compressed representation of~\eqref{eq:u expansion} in an analytically and numerically more favourable format by exploiting an assumed low-rank structure.
Hierarchical representations have appealing properties making them attractive for the treatment of the problems at hand.
For example, they contain sparse polynomials, but are much more flexible at a price of a slightly larger overhead, see e.g. \cite{bachmayr2018parametric,bachmayr2016adaptive} for a comparison concerning parametric PDEs. 

To introduce the concept of \textit{subspace approximations}, which is central to the complexity properties of tensor formats, we start with the classical \textit{Tucker format}.
Given a tensor $\mathbf U$ and a \emph{rank tuple} $\mathbf r:=(r_j)_{j=1}^d$, the approximation problem reads: find optimal subspaces $V_j\subset U_j$ such that
\begin{equation}
    \min_{\mathbf V\in\mathcal V_d} \|\mathbf U - \mathbf V\|\qquad\text{with}\quad \mathcal V_d := \bigotimes_{j=1}^d V_j 
\end{equation}
is minimized over $V_1,\ldots,V_d$, with $\dim V_j=r_j$.
An equivalent problem is to find the corresponding basis vectors $\{b^j_{k_j}\}_{k_j=1,\ldots,r}$ of $V_j$ which can be written in the form
\begin{equation}
    \label{eq:basis Ud}
    b^j_{k_j} := \sum_{\nu_j=1}^{p_j} b^j(\nu_j,k_j)\mathbf e_{\nu_j}^j,\qquad k_j=1,\ldots,r_j<p_j.
\end{equation}
Note that this can be understood as the construction of a reduced basis.
The optimal tensor $\mathbf{V}$ can thus be represented by
\begin{equation}
    \label{eq:Tucker}
    \mathbf V = \sum_{k_1}^{r_1}\cdots\sum_{k_d=1}^{r_d} \mathbf c(k_1,\ldots,k_d)b^1_{k_1}\otimes\cdots\otimes b^d_{k_d} \in \mathcal{V}_d.
\end{equation}
In case of orthonormal bases $\{b^j_{k_j}\}_{k_j=1,\ldots,r_j}$, the \textit{core tensor} $\mathbf c\in\bigotimes_{j=1}^d\mathbb R^{p_j}$ is given entry-wise by projection,
\begin{equation}
    \mathbf c(k_1,\ldots,k_d) = (\mathbf v,b^1_{k_1}\otimes\cdots\otimes b^d_{k_d}).
\end{equation}
With a complexity of $\mathcal O(p_jr_j)$ for each basis $\{b^j_{k_j}\}_{k_j=1,\ldots, r_j}$ and a complexity of $\mathcal O(r^d)$ for the core tensor $\mathbf{c}$, the complexity of the Tucker representation~\eqref{eq:Tucker} is $\mathcal O(pdr + r^d)$ with $r:=\max\{r_j:\; j=1,\ldots,d\}$ and $p:=\max\{p_j:\; j=1,\ldots,d\}$.
As such, the Tucker representation is not sufficient to cope with exponential representation complexity and the format exhibits other problems such as non-closedness.
Nevertheless, the ideas described above eventually lead to a very efficient format by hierarchization of the bases as described in what follows.

\subsection{Hierarchical tensor representations}
\label{sec:HT}

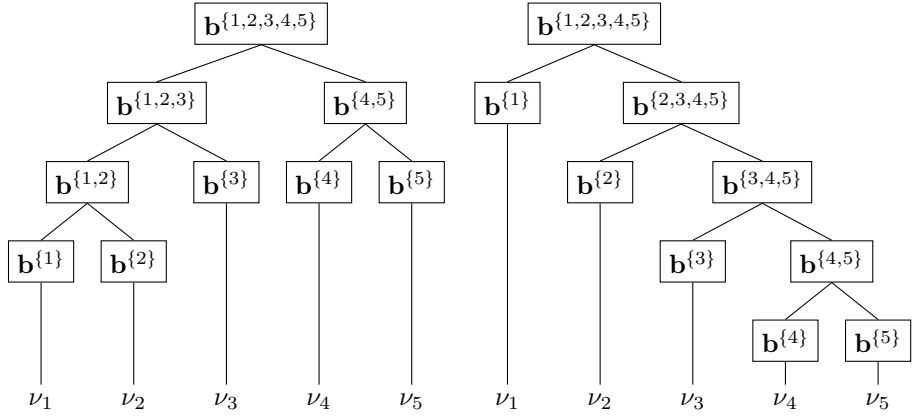
\begin{figure}[!ht]
    \begin{tikzpicture}[sibling distance=10pt]
        \tikzset{frontier/.style={distance from root=140pt}}
        \Tree [.\node[draw]{$\mathbf b^{\{1,2,3,4,5\}}$};
                [.\node[draw]{$\mathbf b^{\{1,2,3\}}$};
                    [.\node[draw]{$\mathbf b^{\{1,2\}}$};
                        [.\node[draw]{$\mathbf b^{\{1\}}$}; $\nu_1$ ]
                        [.\node[draw]{$\mathbf b^{\{2\}}$}; $\nu_2$ ] ]
                    [.\node[draw]{$\mathbf b^{\{3\}}$}; $\nu_3$ ] ]
                [.\node[draw]{$\mathbf b^{\{4,5\}}$};
                    [.\node[draw]{$\mathbf b^{\{4\}}$}; $\nu_4$ ]
                    [.\node[draw]{$\mathbf b^{\{5\}}$}; $\nu_5$ ] ] ]
    \end{tikzpicture}
    \hfill
    \begin{tikzpicture}[sibling distance=10pt]
        \tikzset{frontier/.style={distance from root=140pt}}
        \Tree [.\node[draw]{$\mathbf b^{\{1,2,3,4,5\}}$};
                [.\node[draw]{$\mathbf b^{\{1\}}$}; $\nu_1$ ]
                [.\node[draw]{$\mathbf b^{\{2,3,4,5\}}$};
                    [.\node[draw]{$\mathbf b^{\{2\}}$}; $\nu_2$ ]
                    [.\node[draw]{$\mathbf b^{\{3,4,5\}}$};
                        [.\node[draw]{$\mathbf b^{\{3\}}$}; $\nu_3$ ]
                        [.\node[draw]{$\mathbf b^{\{4,5\}}$};
                        [.\node[draw]{$\mathbf b^{\{4\}}$}; $\nu_4$ ]
                        [.\node[draw]{$\mathbf b^{\{5\}}$}; $\nu_5$ ] ] ] ] ]
    \end{tikzpicture}
    \label{fig:dimension tree}
    \caption{Dimension trees $\mathbb D$ for $d=5$. Balanced HT tree (left) and linearized TT tree (right).}
\end{figure}

The \textit{hierarchical Tucker} (HT) format introduced in~\cite{Hackbusch-2010} is an extension of the notion of subspace approximation to a hierarchical setting determined by a dimension tree as shown in Figure~\ref{fig:dimension tree} where the indices $j=1,\ldots,d$ correspond to the spaces $U_j$ of the tensor space $\mathcal H_d$.
Note that by cutting any edge in the tree, two subtrees are generated.
Collecting the indices for each subtree, a tensor of order two (a matrix) arises.
By this, fundamental principles from matrix analysis, in particular the singular value decomposition (SVD), can be transferred to the higher-order tensor setting.

To illustrate the central idea, consider the optimal Tucker-subspaces $V_1\otimes V_2 \subseteq U_1\otimes U_2=\mathbb R^{p_1}\otimes\mathbb R^{p_2}$.
For the approximation of $\mathbf u\in\mathcal H_d$, often only a subspace $V_{\{1,2\}}\subset V_1\otimes V_2$ with dimension $\operatorname{dim}(V_{\{1,2\}}) = r_{\{1,2\}} < r_1r_2 = \operatorname{dim}(V_1\otimes V_2)$ is required.
In fact, $V_{\{1,2\}}$ is defined by a basis
\begin{equation}
    V_{\{1,2\}} = \mathrm{span}\left\{b^{\{1,2\}}_{k_{\{1,2\}}}:\; k_{\{1,2\}}=1,\ldots,r_{\{1,2\}}\right\}
\end{equation}
with basis vectors
\begin{equation}
    b^{\{1,2\}}_{k_{\{1,2\}}} = \sum_{k_1=1}^{r_1}\sum_{k_2=1}^{r_2} \mathbf b^{\{1,2\}}(k_1,k_2,k_{\{1,2\}})b^1_{k_1}\otimes b^2_{k_2},\quad k_{\{1,2\}}=1,\ldots,r_{\{1,2\}}
\end{equation}
and coefficient tensors $\mathbf b^{\{1,2\}}\in\mathbb R^{r_1\times r_2\times r_{\{1,2\}}}$ where
\begin{equation}
    b^{\{j\}}_{k_{\{j\}}} := \sum_{\nu_j=1}^{p_j} \mathbf{b}^{\{j\}}(\nu_j,k_{\{j\}})\mathbf e_{\nu_j}^j,\qquad j=1,2\text{ and }k_{\{j\}}=1,\ldots,r_j<p_j.
\end{equation}
are the basis vectors of the Tucker representation~\eqref{eq:basis Ud}.
This can be generalized to the tensor product space $\mathcal H_d$ by the introduction of a \textit{partition tree} (or \textit{dimension tree}) $\mathbb D$ with vertices $\alpha\subset D:=\{1,\ldots,d\}$ and leaves $\{1\},\ldots,\{d\}$ where $D$ is called the root of the tree.
Each vertex $\alpha$ that is not a leaf can be partitioned as $\alpha=\alpha_1\cup\alpha_2$ with $\alpha_1\cap\alpha_2=\emptyset$ and $\alpha_1,\alpha_2\ne\emptyset$.
Although not required, one we restrict the topology to a binary tree  %
and denote by $\alpha_1, \alpha_2$ the children of $\alpha$.
Figure~\ref{fig:dimension tree} is an illustration of the unbalanced tree $\mathbb D=\left\{ \{1\},\{2\},\{1,2\},\{3\},\{1,2,3\},\ldots,\{d\},\{1,\ldots,d\} \right\}$ where e.g. $\alpha=\{1,2,3\}=\alpha_1\cup\alpha_2=\{1,2\}\cup\{3\}$.

Let $\alpha_1,\alpha_2\subset D$ be the two children of $\alpha\in D$.
Then $V_\alpha\subset V_{\alpha_1}\otimes V_{\alpha_2}$ is defined by a basis
\begin{equation}
    \label{eq:HT basis}
    b^\alpha_\ell = \sum_{i=1}^{r_{\alpha_1}}\sum_{j=1}^{r_{\alpha_2}} \mathbf b^\alpha(i,j,\ell)b^{\alpha_1}_i\otimes b^{\alpha_2}_j,
\end{equation}
where the tensors $(i,j,\ell)\mapsto \mathbf b^\alpha(i,j,\ell)$ are called \textit{transfer} or \textit{components tensors} and $\mathbf b^D = \mathbf b^{\{1,\ldots,d\}}$ is called the \textit{root tensor}.
To represent a tensor in this hierarchical format it suffices to store the transfer tensors $\mathbf b^\alpha$ along with the root tensor $\mathbf{b}^D$.
More specifically, $\mathbf u\in\mathcal H_d$ is obtained from $(\mathbf b^\alpha)_{\alpha\in\mathbb D}$, via the multilinear function $\tau$
\begin{equation}
    (\mathbf b^\alpha)_{\alpha\in\mathbb D} \mapsto \mathbf u = \tau(\{\mathbf b^\alpha:\;\alpha\in\mathbb D\}),
\end{equation}
which is defined by the recursive application of the basis representation~\eqref{eq:HT basis}.
The mapping $\tau$ is a multilinear function in its arguments $\mathbf b^\alpha$.
A graphical representation of this mapping is depicted in Figure~\ref{fig:dimension tree}.
In this pictorial description, the contractions of component tensors~\eqref{eq:HT basis} are indicated as edges between vertices of a graph and the indices of the tensor are represented by open edges.
This hierarchical representation has complexity $\mathcal O(pdr + dr^3)$ with $p=\max\{p_1,\ldots,p_d\}$ and $r=\max\{r_\alpha:\;\alpha\in\mathbb D\}$.

\paragraph{Tensor trains}
Tensor trains are a subset of the general hierarchical tensors described above.
They were introduced to the numerical mathematics community in~\cite{Oseledets2009,oseledets2010tt} but have been known to physicists for a long time as matrix product states (MPS).
The linear structure is depicted in Figure~\ref{fig:dimension tree} (right), which corresponds to taking $V_{1,\ldots,j+1}\subset V_{\{1,\ldots,j\}}\otimes V_{\{j+1\}}$.
In the example, we consider the unbalanced tree $\mathbb D=\left\{ \{1\},\{2\},\{1,2\},\{3\},\{1,2,3\},\ldots,\{d\},\{1,\ldots,d\} \right\}$.
Applying the recursive construction, any tensor $\mathbf u\in\mathcal H_d$ can be written as
\begin{align}
    (\nu_1,\ldots,\nu_d) &\mapsto \mathbf U(\nu_1,\ldots,\nu_d)\notag\\
    &= \sum_{k_0}^{r_0}\cdots\sum_{k_d}^{r_d} \mathbf U^1(k_0,\nu_1,k_1)\mathbf U^2(k_1,\nu_2,k_2)\cdots \mathbf U^d(k_{d-1},\nu_d,k_d), \label{eq:the train}
\end{align}
where %
\begin{align*}
    \mathbf{U}^1(\nu_1,k_1) &:= \sum_{\ell=1}^{r_{1}} \mathbf{b}^{\{1\}}(\nu_1, \ell) \mathbf{b}^{D}(k_{1},\ell) , \\
    \mathbf{U}^j(k_{j-1},\nu_j,k_j) &:= \sum_{\ell=1}^{r_{j}} \mathbf{b}^{\{j\}}(\nu_j, \ell) \mathbf{b}^{\{j,\ldots,d\}}(k_{j-1},k_j,\ell) ,  \qquad j=2,\ldots,d-1 \\
    \mathbf{U}^d(k_{d-1},\nu_d) &:= \mathbf{b}^{\{d\}}(\nu_d, k_{d-1}) .
\end{align*}
This can be reformulated as matrix products
\begin{equation}
    \label{eq:TT}
    \mathbf U(\nu_1,\ldots,\nu_d) = \prod_{j=1}^d \mathbf b_j(\nu_j) = \tau(\mathbf b^1, \ldots, \mathbf b^d)(\nu),
\end{equation}
with component matrices $b_j(\nu_j)\in\mathbb R^{r_{j-1}\times r_j}$ given by
\begin{equation}
    \left(b_j(\nu_j)\right)_{k_{j-1},k_j} = \mathbf b^j(k_{i-j},\nu_j,k_j),\quad 1< j< d,
\end{equation}
and
\begin{equation}
    \left(b_1(\nu_1)\right)_{k_1}^\intercal = \mathbf b^1(\nu_1,k_1),\quad \left(b_d(\nu_d)\right)_{k_d} = \mathbf b^d(k_d,\nu_d).
\end{equation}
It has to be pointed out that the representation~\eqref{eq:TT} is not unique since in general there exist $\mathbf b^\alpha\neq \mathbf c^\alpha$ such that $\tau\left(\{\mathbf b^\alpha:\; \alpha\in\mathbb D\}\right)=\tau\left(\{\mathbf c^\alpha:\; \alpha\in\mathbb D\}\right)$.
This can also be seen easily in~\eqref{eq:TT} when introducing arbitrary orthogonal matrices and their respective inverses in between the component tensors.

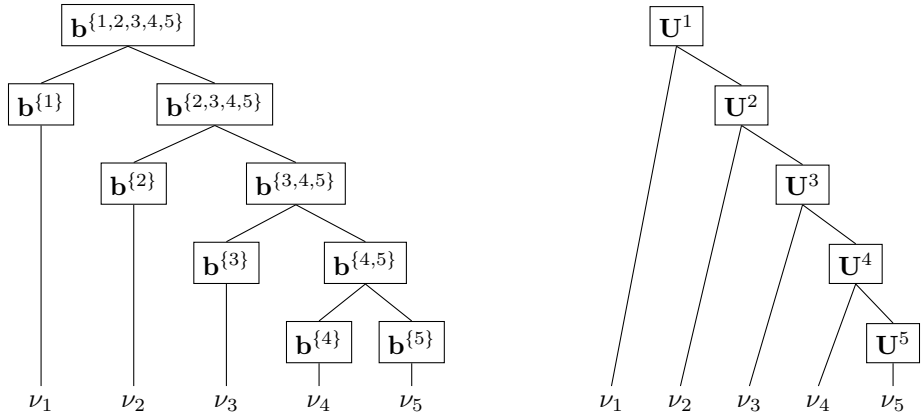
\begin{figure}[!ht]
    \centering
       \begin{tikzpicture}[sibling distance=10pt]
        \tikzset{frontier/.style={distance from root=140pt}}
        \Tree [.\node[draw]{$\mathbf b^{\{1,2,3,4,5\}}$};
                [.\node[draw]{$\mathbf b^{\{1\}}$}; $\nu_1$ ]
                [.\node[draw]{$\mathbf b^{\{2,3,4,5\}}$};
                    [.\node[draw]{$\mathbf b^{\{2\}}$}; $\nu_2$ ]
                    [.\node[draw]{$\mathbf b^{\{3,4,5\}}$};
                        [.\node[draw]{$\mathbf b^{\{3\}}$}; $\nu_3$ ]
                        [.\node[draw]{$\mathbf b^{\{4,5\}}$};
                        [.\node[draw]{$\mathbf b^{\{4\}}$}; $\nu_4$ ]
                        [.\node[draw]{$\mathbf b^{\{5\}}$}; $\nu_5$ ] ] ] ] ]
        \end{tikzpicture}
        \hfill
        \begin{tikzpicture}[sibling distance=10pt]
            \tikzset{frontier/.style={distance from root=140pt}}
            \Tree [.\node[draw]{$\mathbf U^{1}$}; $\nu_1$
                    [.\node[draw]{$\mathbf U^{2}$}; $\nu_2$
                        [.\node[draw]{$\mathbf U^{3}$}; $\nu_3$
                            [.\node[draw]{$\mathbf U^{4}$}; $\nu_4$
                                [.\node[draw]{$\mathbf U^{5}$}; $\nu_5$ ] ] ] ] ]
        \end{tikzpicture}
    \caption{An order $5$ tensor in tensor train representation and its linear representation using component tensors as in~\eqref{eq:the train}.}
    \label{TT:fig:hosvd}
\end{figure}

An illustration of the tensor train structure~\eqref{eq:the train} is depicted in Figure~\ref{TT:fig:hosvd} (right), which is equivalent to the tree structure shown on the left-hand side.

It turns out that every tensor has a TT-representation with minimal rank, which means that the TT-rank is well-defined.
Moreover, an efficient algorithm for computing a minimal TT-representation is given by the TT Singular Value Decomposition (TT-SVD)~\cite{holtz2012manifolds}.
Additionally, the set of tensor trains with fixed TT-rank $\mathbf r$ denoted by $\mathcal T_\mathbf r\subseteq \mathcal H_d$ forms a smooth manifold.
If all lower ranks are included, an algebraic variety denoted by $\mathcal T_{\leq\mathbf r}$ is formed \cite{kutschan2017tangent}.

\subsection{Tensor Trains as  differentiable manifolds}
\label{sec:differentiable}

The multilinear structure of the tensor product enables efficient optimization within the manifold structure.
Endowed with the Euclidean metric induced by the Frobenius scalar product, the set $\mcal{T}_{\mathbf r}$ becomes an embedded Riemannian manifold~\cite{holtz2011TTMfld,uschmajew2020riemannianTT,wolf2019}.
This allows the formulation of different line search algorithms utilizing the \emph{Riemannian gradient}.
For a function $J : \mcal{H}_n\to \mbb{R}$ the Riemannian gradient at $X\in\mcal{T}_r$ can be computed by projecting the Euclidean gradient onto the tangent space $\mathbb{T}_X$ at $X$ (see e.g.~\cite{steinlechner2016thesis,absil2008book}), i.e.
\begin{equation}
    P_{\mathbb{T}_{X}} \nabla J\pars{X},
\end{equation}
where $P_{\mathbb{T}_{\widetilde{X}}}$ is the projector onto the tangent space of $\mcal{T}_{r}$ at the point $\widetilde{X}$.
Just as the negative Euclidean gradient, the negative Riemannian gradient can be used as a descent direction for minimizing $V_{p,N}^{m}$.
In theory, the strategy is to move in that direction along a geodesic until a local minimum is reached.
Starting from $\widetilde{X}$, the function that moves in the direction $Z\in\mbb{T}_{\widetilde{X}}$ along a geodesic for a distance of $\norm{Z}$ is called the \emph{exponential map} $\exp_{\widetilde{X}}\pars{Z}$.
Unfortunately, there is no analytic expression for the exponential map available for $\mcal{T}_{r}$.
Instead, one usually resorts to a so-called \emph{retraction} $\mcal{R}_{\widetilde{X}}\pars{Z}$ which is an approximation of the exponential map, see~\cite{absil2008book} for details.
In the tensor train format, an example of a retraction is defined by the TT-SVD via
\begin{equation}
    \mcal{R}_{\widetilde{X}}\pars{Z} = \operatorname{TT-SVD}\pars{\widetilde{X} + Z}
\end{equation}
as shown by~\cite{steinlechner2016thesis}.
Using these techniques, a steepest descent update with step size $\beta$ on the manifold $\mcal{T}_{r}$ is given by
\begin{equation}
    \widetilde{X}_{k+1} = \mcal{R}_{\widetilde{X}_k}\pars{- \beta P_{\mbb{T}_{\widetilde{X}_k}} \nabla V_{p,N}^m\pars{\widetilde{X}_k} } .
\end{equation}

Convergence of Riemannian optimization algorithms is typically only considered for smooth functions.
When this can be assumend, the convergence can be sped up by using higher-order algorithms such as the conjugated gradient method.
This additionally requires a method of ``moving'' tangent vectors $Z_{k-1}\in\mbb{T}_{\widetilde{X}_{k-1}}$ from the tangent space at point $\widetilde{X}_{k-1}$ to the tangent space $\mbb{T}_{\widetilde{X}_{k}}$ at point $\widetilde{X}_{k}$.
Again, the optimal differential geometric tool, the \emph{parallel transport}, is computationally infeasible on the tensor train manifold.
However, the \emph{vector transport} introduced by \cite{absil2008book} defines a class of approximations, which can be used to accomplish this task.
In the tensor train format, such a vector transport is given by the projection $P_{\mbb{T}_{\widetilde{X}_{k}}} Z_{k-1}$.

\section{A version of the Longstaff-Schwartz algorithm based on the Tensor Train format}
\label{sec:primal minimization}

We now combine the tensor train format introduced in Section~\ref{sec:HT} with the Longstaff-Schwartz algorithm for computing Bermudan option prices as detailed in Algorithm~\ref{alg:LS}.
To make the approximation problem~\eqref{eq:cond_exp_approx} concrete a set of basis functions $\{\vec{B}_\alpha\}_{\alpha\in\Lambda}$ has to be chosen.
We prefer to work on a compact sub-domain of the reals, which we choose such that the probability of assets lying outside the domain is minimal.
As a heuristic method for determining the truncation, we set
\[
    a = \min_{m,n,k} (S^m_n)_k \qquad\text{and}\qquad b = \max_{m,n,k} (S^m_n)_k
\]
and choose the $H^2(a, b)$-orthogonal basis functions $\{ B_1, \dots, B_{p} \}$ spanning the space of polynomials of degree $p$.
We then represent the approximation of the discounted value of the option $v:\mathbb R^{d'} \to \mathbb R$ by
\begin{equation}
\label{eq:approximate value}
    v(x) = \sum_{\alpha\in\Lambda} V_\alpha \prod_{k=1}^{d'} B_{\alpha_k}(x_k),
\end{equation}
where we approximate the coefficient tensor $\mathbf V\in(\mathbb{R}^{p})^{\otimes d'}$ in the TT format.
As is common practice in Longstaff-Schwartz type algorithms we augment this basis by the payoff function $\varphi$.
With the definition
\[
    B: \mathbb R \to \mathbb R^p,\quad B(x) = [B_1(x), \dots, B_p(x) ],
\]
i.e. $B$ stacks the one-dimensional basis functions into a vector such that they can be contracted with the component tensors,
the resulting approximation $v:\mathbb R^{d'} \to \mathbb R$ is graphically represented by
\begin{center}
    \begin{tikzpicture}
        \begin{scope}[every node/.style={draw,  fill=white}]
            \node (A1) at (0,0) {$U_1$}; 
            \node (A2) at (1.25,0) {$U_2$}; 
            \node (A3) at (2.5,0) {$U_3$}; 
            \node (A4) at (4.25,0) {$U_{d'}$}; 
            
            \node (B1) at (0,-1) {$B(x_1)$}; 
            \node (B2) at (1.25,-1) {$B(x_2)$}; 
            \node (B3) at (2.5,-1) {$B(x_3)$}; 
            \node (B4) at (4.25,-1) {$B(x_{d'})$}; 
        \end{scope}
        \node (C0) at (-2,0) {$v(x)$}; 
        \node (C1) at (-1,0) {$=$}; 
        \node[right=0.25 of A4] (plus) {$+$}; 
        \node[right=0.05 of plus] (g) {$c_\varphi \varphi(x)\ .$};
        \begin{scope}[every edge/.style={draw=black,thick}]
        	\path [-] (A1) edge node[midway,left,sloped] [above] {$r_1$} (A2);
        	\path [-] (A2) edge node[midway,left,sloped] [above] {$r_2$} (A3);
        	\path [-] (A1) edge node[midway,left] [right] {$p$} (B1);
        	\path [-] (A2) edge node[midway,left] [right] {$p$} (B2);
        	\path [-] (A3) edge node[midway,left] [right] {$p$} (B3);
        	\path [-] (A4) edge node[midway,left] [right] {$p$} (B4);
        	\path [-] (A3) edge ($(A3)+(0.56,0)$);
        	\draw[cheating dash=on 2pt off 2pt, thick] ($(A3)+(0.55,0)$) edge ($(A4)-(0.55,0)$);
        	\path [-] ($(A4)-(0.56,0)$) edge (A4);
        \end{scope}
    \end{tikzpicture}
\end{center}
Note that on the r.h.s.\ of this equation every open-index of $U_i$ and $B(x_i)$ for $1 \leq i \leq d'$ is contracted, which indeed results in a scalar value $v(x)$.

To solve the resulting minimization problem~\eqref{eq:cond_exp_approx} we use a rank adaptive version of the \emph{alternating least-squares (ALS)} algorithm \cite{ALS}, the \emph{stable alternating least-squares algorithm (SALSA)} \cite{grasedyck2019stable}.
Using this algorithm relieves us from having to guess an appropriate rank of the solution beforehand.
As a termination condition we check whether the error on the samples or on a validation set decreases sufficiently during one iteration.
In our implementation this validation set is chosen to have $20\%$ of the size of the training set.

We now describe how we modify ALS (or SALSA) to handle the additional term $c_\varphi \varphi(x)$.
The classical ALS algorithm optimizes the component tensors $\{U_1,\ldots,U_{d'}\}$ in an alternating fashion.
For each $k=1,\ldots,d'$ all component tensors $\{U_j\}_{j\ne k}$ are fixed and only $U_k$ is optimized.
This procedure is then repeated alternatingly until a convergence criterion is met.

We modify this scheme by optimizing $c_\varphi$ as well as $U_k$ for each $k$.
Since the mapping $(U_k,c_\varphi) \mapsto v$ is linear, the resulting problem is a classical linear least squares problem
\[
    (U_k, c_\varphi) = \argmin_{w,c} \frac{1}{m}\sum_{i=1}^{m} |Y^m - A_k^m(w,c)|^2 .
\]
To exemplify this, for $k=2$ the operator $A_k^m$ is diagrammatically represented by
\begin{center}
    \begin{tikzpicture}
        \begin{scope}[every node/.style={draw,  fill=white}]
            \node (A1) at (0.0,0) {$U_1$}; 
            \node (A2) at (1.5,0) {$w$}; 
            \node (A3) at (3.0,0) {$U_3$}; 
            \node (A4) at (5.0,0) {$U_{d'}$}; 
            
            \node (B1) at (0.0,-1) {$B(S_1^m)$}; 
            \node (B2) at (1.5,-1) {$B(S_2^m)$}; 
            \node (B3) at (3.0,-1) {$B(S_3^m)$}; 
            \node (B4) at (5.0,-1) {$B(S_{d'}^m)$}; 
        \end{scope}
        \node (C0) at (-2.5,0) {$A_k^m (w, c_\varphi)$}; 
        \node (C1) at (-1,0) {$=$}; 
        \begin{scope}[every edge/.style={draw=black,thick}]
        	\path [-] (A1) edge node[midway,left,sloped] [above] {$r_1$} (A2);
        	\path [-] (A2) edge node[midway,left,sloped] [above] {$r_2$} (A3);
        	\path [-] (A1) edge node[midway,left] [right] {$p$} (B1);
        	\path [-] (A2) edge node[midway,left] [right] {$p$} (B2);
        	\path [-] (A3) edge node[midway,left] [right] {$p$} (B3);
        	\path [-] (A4) edge node[midway,left] [right] {$p$} (B4);
        	\path [-] (A3) edge ($(A3)+(0.56,0)$);
        	\draw[cheating dash=on 2pt off 2pt, thick] ($(A3)+(0.55,0)$) edge ($(A4)-(0.55,0)$);
        	\path [-] ($(A4)-(0.56,0)$) edge (A4);
        \end{scope}
        \node[right=0.25 of A4] (plus) {$+$}; 
        \node[right=0.05 of plus] (g) {$c_\varphi \varphi(S^m)\ .$};
    \end{tikzpicture}
\end{center}
After reshaping the pair $(w,c)\in\mathbb{R}^{r_1\times p\times r_2}\times\mathbb{R}$ into a vector of size $r_1pr_2+1$, the operator can be written as $A\in\mathbb{R}^{m\times (r_1p_2r_2+1)}$ and the problem becomes
\[
    X = \argmin_{x} \frac{1}{m} \|\vec{Y} - Ax\|_2^2,
\]
where $\vec{Y} = [Y^1,\ldots,Y^m]$ .

\subsection*{Complexity analysis}
\label{sec:primal complexity}

Using a tensor train representation instead of the full tensor allows us to reduce the space complexity from $\mcal{O}\pars{p^{d'}}$ to $\mcal{O}\pars{d'pr^2}$ with $r = \max\braces{r_1,\ldots,r_{d'-1}}$.
For moderate $r$ this leads to a dramatic reduction in memory usage which we observe in our experiments.
Figure~\ref{fig:ranks_barplot} shows that the rank-adaptive algorithm computes solutions with $r<6$ and we numerically verify that for $d'>100$ a rank of $r=1$ is sufficient for obtaining values within the reference interval from the literature.
This allows us to compute the price of max-call options with up to $1000$ assets.

Since ALS is an iterative method its time complexity can only be provided per iteration and amounts to
\begin{equation}
    \mcal{O}\pars{Nm\abs{\Lambda_p}^2r^4}
\end{equation}
floating point operations per iteration.
As with every iterative algorithm the number of iterations needed depends on the specific problem. 
In our numerical tests we generally needed less than $10$ iterations.

\section{Dual martingale minimization with tensor trains}
\label{sec:dual minimization}

To use the tensor train format in the dual formulation, we define the set $\mcal{P}_{\hat{0}} = \braces{\widetilde{X} : \widetilde{X}_0 = 0}$ and rewrite \eqref{eq:dual_disc} as
\begin{equation}
    U_0 = \inf_{\substack{\widetilde{X} \in \mcal{T}_{r} \cap \mcal{P}_{\hat{0}}}} V_{p,N}^m\pars{\widetilde{X}}, \label{eq:dual_disc_intersection}
\end{equation}
where $\mathcal{T}_{r}$ denotes the set of TT tensors of rank $r$ and $V_{p,N}^m$ is the cost function that is minimized in \eqref{eq:dual_disc}.
Performing this optimization directly on the parameters of the tensor train is ill-posed since its parametrization is not unique.
A common way to solve this is to use the manifold structure of $\mcal{T}_r$ and employ a Riemannian optimization algorithm.
For this~\eqref{eq:dual_disc_intersection} has to be rephrased as an unconstrained smooth optimization problem.

Define the projector $\pars{P_{\hat{0}} \widetilde{X}}_{{\alpha}} = (1-\delta_{{\alpha}0}) \widetilde{X}_\alpha$ and remove the constraint $\widetilde{X}\in\mcal{P}_0$ by rewriting \eqref{eq:dual_disc_intersection} as
\begin{equation}
    U_0 = \inf_{\widetilde{X} \in \mcal{T}_{r}} V_{p,N}^m\pars{P_{\hat{0}} \widetilde{X}} .
\end{equation}
Since $P_{\hat{0}}$ is a linear operator, the modified cost function $V_{p,N}^m\circ P_{\hat{0}}$ retains the convexity, continuity and piece-wise linearity of $V_{p,N}^m$.
We then mollify the $V_{p,N}^m$ by replacing the maximum with the smooth approximation
\begin{equation}
    \amax_{n=1,\ldots,N} x_n = \frac{\sum_{n=1}^N x_n e^{\alpha x_n}}{\sum_{n=1}^N e^{\alpha x_n}} .
\end{equation}
The resulting cost function reads
\begin{equation}
    V_{p,N}\pars{\widetilde{X}} = V_{p,N}^{m,\alpha}\pars{\widetilde{X}} = \frac{1}{m}\sum_{i=1}^m \bracs*{\amax_{n=1,\ldots,N} \pars*{Z^{\pars{i}}_{t_n} - \sum_{\alpha\in\Lambda^n} \widetilde{X}_\alpha H_{\alpha_1}\pars{G_1^{\pars{i}}} \cdots H_{\alpha_n}\pars{G_n^{\pars{i}}}}}.
\end{equation}
The respective optimization problem
\begin{equation}
\label{eq:U0 optimization}
    U_0 = \inf_{\widetilde{X} \in \mcal{T}_{r}} V_{p,N}^{m,\alpha}\pars{P_{\hat{0}} \widetilde{X}}
\end{equation}
can be solved by Riemannian algorithms.
We use a conjugated gradient method with the \textsc{FR-PR\textsubscript{+}} update rule as defined in~\cite{cg}.

We also have to address the choice of the initial value for the optimization.
Since the set $\mcal{T}_{r}$ is not convex, a diligent choice is important in order to reach the global minimum.
We obtain such a value for polynomial degree $p$ by using the optimal value $\widetilde{X}^{\pars{p-1}}$ for the polynomial degree $p-1$.
This recursion stops at $p=0$ where we know the optimal value to be $\widetilde{X}^{\pars{0}} = 0$.

In our implementation we used a constant rank of $4$ and chose $\alpha = 50$ which, empirically, held the smoothing induced error below $10^{-3}$.
As a termination condition we check if the error does not sufficiently decrease over a period of $10$ iterations.
Of all iterates obtained during the optimization we choose the one that has the lowest value on a validation set.
In our implementation this validation set is chosen to have one ninth of the size of the training set.

\subsection*{Complexity analysis}
\label{sec:dual complexity}

In the dual method we observe the same dramatic reduction in space complexity as in the primal algorithm.
The space complexity of $\mcal{O}\pars{p^{Nd'}}$ for the full tensor is reduced to $\mcal{O}\pars{Nd'pr^2}$ for a tensor in the tensor train format with a rank uniformly bounded by $r$.
This allows us to use the dual algorithm to compute the price of a basket put option with $N=31$ exercise dates in Table~\ref{tbl:experiment_putbasket_lelong}.

Since gradient descent is again an iterative algorithm the time complexity can only be computed per iteration.
Assuming that $\widetilde{X}$ is a tensor train tensor with rank $r$, the contraction
\begin{equation}
    \sum_{\alpha\in\Lambda} \widetilde{X}_\alpha H_{\alpha_1}\pars{G_1^{\pars{i}}} \cdots H_{\alpha_n}\pars{G_n^{\pars{i}}}
\end{equation}
can be computed with $\mcal{O}\pars{n\abs{\Lambda_p}r^2 + \pars{N-n}r^2}$ floating point operations.
This means that both $V_{p,N}^{m,\alpha}\pars{\widetilde{X}}$ and its gradient can be computed with $\mcal{O}\pars{m N^2 \abs{\Lambda_p}r^2}$ floating point operations.
Compare this to the $\mcal{O}\pars{mp^{Nd'}}$ floating point operations required for the full tensor and to the $\binom{Nd'+p}{Nd'}$ operations for the sparse tensor.
At least from a theoretical point of view, evaluation and optimization are faster in the tensor train format, namely
\begin{itemize}
    \item exponentially faster when compared to the full tensor ansatz and
    \item when $p>2$ up to a polynomial factor for the sparse ansatz.
\end{itemize}
These statements obviously depend on the rank $r$ which is bounded by at most $4$ in our experiments, meaning that the represented objects are in fact low-rank.

\section{Numerical experiments}
\label{sec:experiments}

In this part, we present results obtained from the algorithms described above. 
Implementations in Python can be found at \url{https://github.com/ptrunschke/tensor_option_pricing}.
For each experiment, we report low-biased estimators $v_0\pars{S_{t_0}}$ and $V_{p,N}^{m,\alpha}\pars{\widetilde{X}}$ based on re-simulated trajectories, see \cite{glasserman2013monte}. More precisely, we generate independent trajectories of the underlying price process $S$ and apply the stopping strategy implied by the already computed approximate value functions $v_k$, giving as a low-biased approximation to the true option price. Conversely, approximately optimal martingale parameterizations computed by the dual algorithm are used to compute a high-biased estimator, once again based on new trajectories, not used to produce the parameterization in the first place.

In the following we denote by $n$ the number of possible exercise dates, including $0$, by $p$ the polynomial degree used in the approximation of the conditional probabilities and in the Wiener--It\^o chaos expansion and by $m$ the number of samples used.
We further denote by $m_{\mathrm{resim}}$ the number of samples used for the resimulation.
$V_{\text{LS}}$ is the price computed by the resimulation of the Longstaff--Schwartz method and $V_{\text{dual}}$ is the price computed by the dual method.
The corresponding reference values are denoted by $V_{\text{LS}}^{\text{ref}}$ and $V_{\text{dual}}^{\text{ref}}$ respectively, and were obtained in the literature -- see specific references for the individual examples.

\subsection{Options in the Black--Scholes model}

The $d$-dimensional Black Scholes model for $j\in \{1, \ldots, d\}$ reads
\begin{equation}\label{eq:black_scholes}
    \mathrm{d}S^j_t = S^j_t(r_t - \delta_t)\mathrm{d}t + \sigma^j L_j\mathrm{d}B_t),
\end{equation}
where $B$ is a Brownian motion with values in $\mathbb{R}^d$, $\sigma = (\sigma_1, \ldots, \sigma_d)$ is the vector of volatilities assumed to be deterministic and positive at all times, and $L_{j}$ is the $j$-th row of the matrix $L$ defined as a square root of the correlation matrix chosen to be of the form
\begin{equation}
    \Gamma = \begin{pmatrix}
        1 & \rho & \cdots & \rho \\
        \rho & 1 & \ddots & \vdots \\
        \vdots & \ddots & \ddots & \rho \\
        \rho & \cdots & \rho & 1 \\
    \end{pmatrix},
\end{equation}
where $\rho \in (-1/(d-1), 1]$ to ensure that $\Gamma$ is positive definite.
The initial condition for the SDE is given by the spot price $S_0$.

We will test the algorithms for different payoff functions $\phi$, dimensions $d$ and strike prices $K$.

\subsection{A basket put option on correlated assets}
We first consider the case of a put basket option on correlated assets.
The payoff of this option writes as $\phi\pars{S_t} = \left(K - \sum_{j=1}^d \omega_jS_t^j\right)_+$ where $\omega = \pars{\omega_1, \ldots, \omega_d}$ is a vector of real valued weights.
We report in Table~\ref{tbl:experiment_putbasket_lelong} and Table~\ref{tbl:experiment_putbasket_lelong_100K} our values compared to the reference prices for two different sample sizes $m = 20\hspace{0.1em}000$ and $m = 10^5$.
Blank cells in the tables indicate that reference values are not reported in the reference papers.
The results of our experiments are reported in Table~\ref{tbl:experiment_putbasket_lelong}.

It can be seen that the values obtained by our version of Lelong's method are not as close to the reference price as are the values obtained by~\cite{lelong2018dual}.
From a theoretical perspective a lower value should always be possible given a sufficient rank.
We thus attribute this to the lack of a rank adaption strategy in the dual problem and highlight this as an interesting direction for further research.
It can moreover be seen that for $N=31$ the values of $V_{\text{dual}}$ increase with $p$.
Because the manifold for $p=2$ is a submanifold of $p=3$ one would expect that this is impossible.
Note however that the table shows resimulated prices only.
Therefore we interpret this observation to indicate that a larger value of $m$ is needed in this case.
This is confirmed in Table~\ref{tbl:experiment_putbasket_lelong_100K}.

For the Longstaff-Schwartz variant we use $m = 10^5$ and observe values close to the reference value. 
Furthermore, in the case $N = 31$ and $S_0^j = 100$,  we observe that the result for $p=2$ dominate the $p=3$ case, indicating sub-optimal results.
However, as seen in Table \ref{tbl:experiment_putbasket_lelong_100K} we obtain better results for polynomial degree $p=8$.
Note that we have capped the TT-rank at $4$ for the computation with $p=8$.
By doing that, the computational time only increased by a factor of $3$ when compared to the run time for the case $p=3$, being $40$ seconds and $15$ seconds respectively.

We also report that during the optimization within the Longstaff-Schwartz algorithm the TT-rank of the value function did not exceed $5$ for any test-case, which means that a low-rank structure of the sought expectation values within the polynomial ansatz space is noticeable.
This low-rank structure is a necessity for high-dimensional computation and will be analyzed in greater detail in the next example.
In this example the number of samples used for training has a larger effect not only on the variances but also on the values.

\begin{table}[!ht]
\centering
\begin{tabular}{ c c c | c c c | c c c }
$p$ & $N$ & $S_0^j$ & $V_{\text{dual}}$ & Stddev & $V_{\text{dual}}^{\text{ref}}$ & $V_{\text{LS}}$ & Stddev & $V^{\text{ref}}$ \\
 \hline
$2$ & $4$  & $100$ & $2.34$ & $0.003$ & $2.29$ & $2.15$ & $0.009$ & $2.17$ \\
$3$ & $4$  & $100$ & $2.33$ & $0.003$ & $2.25$ & $2.16$ & $0.009$ & $2.17$ \\
$2$ & $7$  & $100$ & $2.64$ & $0.002$ & $2.62$ & $2.39$ & $0.008$ & $2.43$ \\
$3$ & $7$  & $100$ & $2.64$ & $0.002$ & $2.52$ & $2.40$ & $0.008$ & $2.43$ \\
$2$ & $31$ & $100$ & $3.08$ & $0.002$ & & $2.49$  & $0.01$ &  \\
$3$ & $31$ & $100$ & $3.12$ & $0.002$ & & $2.36$ & $0.01$ &  \\
\hline
$2$ & $4$  & $110$ & $0.67$ & $0.002$ & $0.57$ & $0.53$ & $0.006$ & $0.55$ \\
$3$ & $4$  & $110$ & $0.67$ & $0.002$ & $0.55$ & $0.53$ & $0.006$ & $0.55$ \\
$2$ & $7$  & $110$ & $0.78$ & $0.002$ & $0.64$ & $0.57$ & $0.007$ & $0.61$ \\
$3$ & $7$  & $110$ & $0.77$ & $0.002$ & $0.64$ & $0.57$ & $0.007$ & $0.61$ \\
$2$ & $31$ & $110$ & $3.94$ & $0.002$ & &  $0.61$ & $0.008$  &  \\
$3$ & $31$ & $110$ & $3.95$ & $0.002$ & & $0.61$ & $0.008$ &  \\
\end{tabular}
\caption{Prices for the put basket option with parameters $d=5$, $T=3$, $r=0.05$, $\delta^j = 0$, $\sigma^j=0.2$, $\rho=0$, $K=100$, $\omega_j=\frac{1}{d}$, $m=20\hspace{0.1em}000$, $m_{\mathrm{resim}}=10^6$. Values for $V_{\text{dual}}^{\text{ref}}$ and $V^{\text{ref}}$ are taken from \cite{lelong2018dual}. Number of samples for Longstaff-Schwartz: $m_{\text{LS}} = 10^5$. Empty spaces denote unavailable reference values.}
\label{tbl:experiment_putbasket_lelong}
\end{table}

\begin{table}[!ht]
\centering
\begin{tabular}[t]{ c c | c c }
$p$ & $S_0^j$ & $V_{\text{dual}}$ & Stddev \\
    \hline
$2$ & $100$ & $2.88$ & $0.001$ \\
$3$ & $100$ & $2.88$ & $0.001$ \\
    \hline
$2$ & $110$ & $0.80$ & $0.001$ \\
$3$ & $110$ & $0.80$ & $0.001$ \\
\end{tabular}
\qquad
\begin{tabular}[t]{ c c | c c }
$p$ & $S_0^j$ & $V_{\text{LS}}$ & Stddev \\
    \hline
$8$ & $100$ & $2.56$ & $0.01$ \\
\end{tabular}
\caption{Prices for the put basket option with parameters $d=5$, $N=31$, $T=3$, $r=0.05$, $\delta^j = 0$, $\sigma^j=0.2$, $\rho=0$, $K=100$, $\omega_j=\frac{1}{d}$, $\mathbf{m=10^5}$, $m_{\mathrm{resim}}=10^6$. Empty spaces denote unavailable reference values}
\label{tbl:experiment_putbasket_lelong_100K}
\end{table}

\FloatBarrier

\subsection{Bermudan max-call options} \label{sec:max-call}

In this section we consider max-call options and in particular the scalability of the tensor train approach for the Longstaff-Schwartz algorithm for higher dimensions.
The reference values for this problem were taken from~\cite{andersen2004primal, becker2019deep}.
The payoff function of a max-call option takes the form
\begin{equation}
    \left( \max_{1 \leq i \leq d} \omega_i S^i - K \right)_+.
\end{equation}
In Table~\ref{tbl:experiment_maxcall_lelong} we report results for the dual algorithm.
In contrast to the case of the put basket option, we see that we are close to the values computed by the original method~\cite{lelong2018dual} and in some cases improve the previously reported results.
This indicates the viability of this approach.
A rank-adaptive algorithm could probably further improve the efficiency of our method in high dimensions.

\begin{table}[h!]
\centering
\begin{tabular}{ c c c c | c c c | c }
$p$ & $d$ & $m$ & $S_0^j$ & $V_{\text{dual}}$ & Stddev & $V_{\text{dual}}^{\text{ref}}$ & $V^{\text{ref}}$ \\
 \hline
$2$ & $2$ & $20\hspace{0.1em}000$ &  $90$ &  $8.85$ & $0.004$ & $10.05$ &  $8.15$ \\
$3$ & $2$ & $20\hspace{0.1em}000$ &  $90$ &  $8.83$ & $0.004$ &  $8.6$  &  $8.15$ \\
$2$ & $5$ & $20\hspace{0.1em}000$ &  $90$ & $21.68$ & $0.014$ & $21.2$  & $16.77$ \\
$3$ & $5$ & $40\hspace{0.1em}000$ &  $90$ & $21.40$ & $0.015$ & $20.13$ & $16.77$ \\
\hline
$2$ & $2$ & $20\hspace{0.1em}000$ & $100$ & $14.68$ & $0.004$ & $16.3$  & $14.01$ \\
$3$ & $2$ & $20\hspace{0.1em}000$ & $100$ & $14.65$ & $0.004$ & $15$    & $14.01$ \\
$2$ & $5$ & $20\hspace{0.1em}000$ & $100$ & $32.37$ & $0.017$ & $31.8$  & $26.34$ \\
$3$ & $5$ & $40\hspace{0.1em}000$ & $100$ & $31.95$ & $0.017$ & $29$    & $26.34$ \\
\end{tabular}
\caption{Prices for the call option on the maximum of $d$ assets with parameters $N=10$, $T=3$, $r=0.05$, $\delta^j = 0.1$, $\sigma^j=0.2$, $\rho=0$, $K=100$, $m_{\mathrm{resim}}=10^6$. Values for $V_{\text{dual}}^{\text{ref}}$ and $V^{\text{ref}}$ are taken from \cite{lelong2018dual}.}
\label{tbl:experiment_maxcall_lelong}
\end{table}

In Table~\ref{tbl:experiment_max_call} we consider the Longstaff-Schwartz algorithm in moderate to extreme dimensions. We increase the number of samples to $10^6$ and test every polynomial degree up to $p=7$.
We observe that we rarely see any significant improvement when using polynomial degree larger than $4$ or $5$.
However, throughout the table polynomial degree $p=6$ appears to obtain the overall best results, with small improvements over the other polynomial degrees.
Moreover, we see that while we are not exactly as high as the reference value for low dimensions, i.e. $d \leq 20$, the results for higher dimension are accurate.
A possible explanation for this is that the value function might have simpler structure in high dimension.

Finally, in Table~\ref{tbl:experiment_max_call_sort} we use a trick, where after sampling all the paths, we sort the assets at every time point by decreasing magnitude, see, e.g., \cite[p. 1230]{andersen2004primal}.
We observe, that , the unsorted algorithm performs better than the sorted, while both stay closely under the reference interval.
We observe, that while the unsorted algorithm is already performing well, sorting the assets yields an increase in performance in every dimension.
Moreover, for the sorted case, polynomial degree of $3$ appears to be sufficient to obtain optimal results.
Finally, we observe some numerical instabilities for our implementation of the sorted algorithm 
when the dimension is $d=750$ or $d=1000$ and the polynomial degree is larger than $3$.
We assume that by using a better polynomial basis these instabilities can be resolved.
However, as polynomial degree $3$ was sufficient in the lower-dimensional case we did not further investigate this instability.
We state that within these experiments the standard deviation of the resimulations was never larger than $0.1$.

It is worth noting, that the results in very high dimensions were obtained by calculating only $10^6$ trajectories while the reference values were computed using more than $24\times 10^6$ paths using state-of-the-art machine learning techniques, see \cite{becker2019deep}.
This underlines the potential of tensor train approaches for optimal stopping, especially in high dimensions.

\begin{figure}[!ht]
    \centering
    \includegraphics[width=\textwidth]{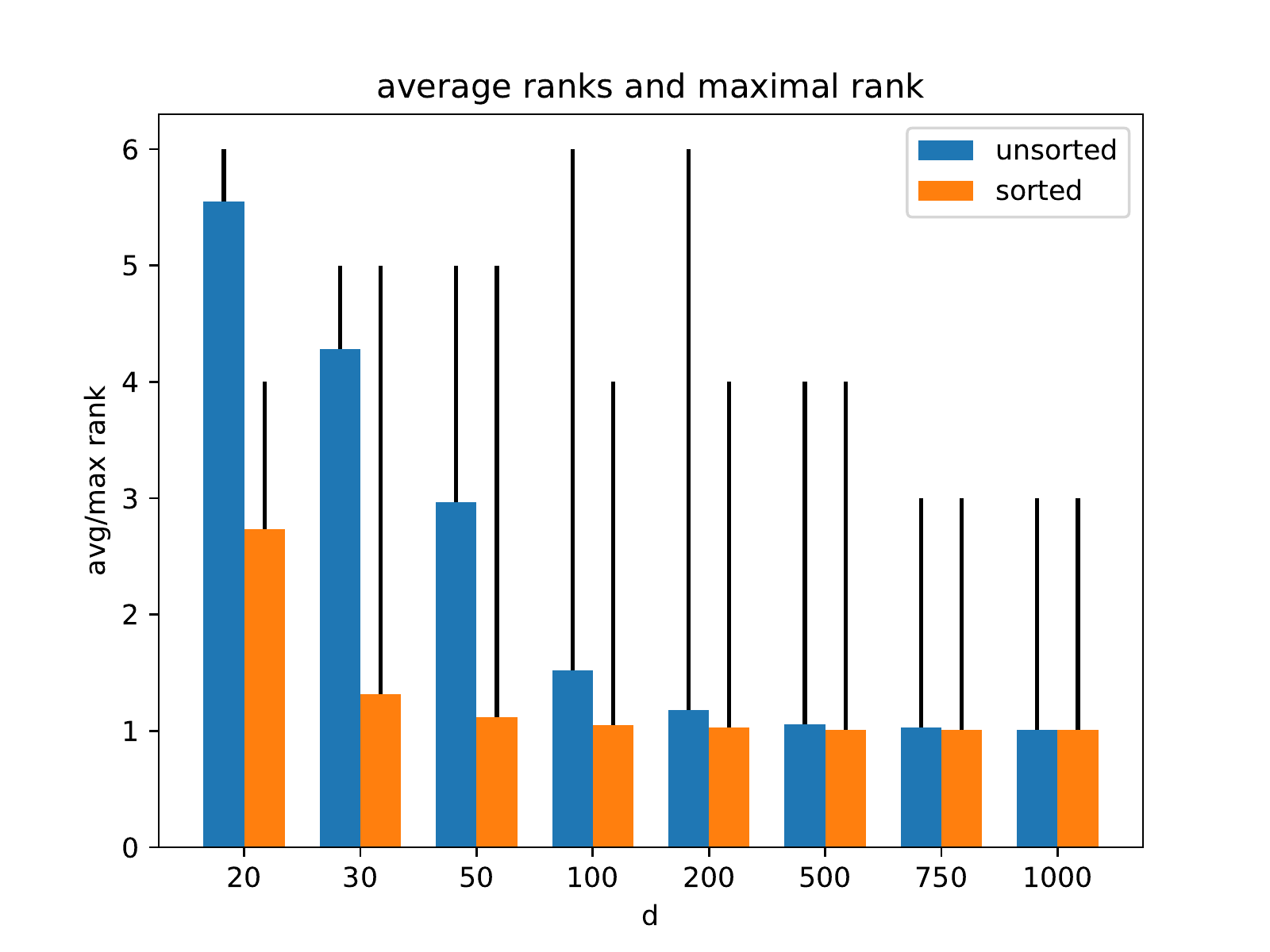}
    \caption{average ranks(blue) and maximal(black) rank for different dimension.
    The black lines indicate the maximal rank.
    We use the results with highest values for every dimension and every bar.}
    \label{fig:ranks_barplot}
\end{figure}

In Figure \ref{fig:ranks_barplot} we analyze the average and the maximal rank of the value function and observe a decrease of the ranks in higher dimensions.
We state that from $d = 100$ a separate test run where we fix the ranks to $1$ yield comparable results, implying that a rank $1$ solution can yield close to optimal results.
This means, that the value function indeed has a simple structure in high dimension.

\begin{table}[h!]
\centering
\begin{tabular}{ c | c c c c c c c | c}
$d$ & \multicolumn{7}{c}{p} & $V_\textrm{ref}$ \\
 \hline
 & $1$ & $2$ & $3$ & $4$ & $5$ & $6$ & $7$ & \\
\hline
$2$ & $13.66$ & $13.79$ & $13.81$ & $13.76$ & $13.80$ & $13.83$ & $13.78$  & $13.902$ \\
$3$ & $18.34$ & $18.30$ & $18.39$ & $18.48$ & $18.50$ & $18.55$ & $18.53$ & $18.69$ \\
$5$ & $25.66$ & $25.58$ & $25.70$ & $25.97$ & $25.75$ & $25.84$  & $25.93$ & $[26.115, 26.164]$\\
$10$ & $37.77$ & $37.65$ & $38.01$ & $38.12$ & $38.25$ & $38.27$ & $38.14$ & $[38.300, 38.367]$ \\
$20$ & $51.10$ & $51.34$ & $51.49$ & $51.64$ & $51.62$ & $51.63$ & $51.62$ & $[51.549, 51.803]$ \\
$30$ & $59.11$ & $59.30$ & $59.50$ & $59.63$ & $59.62$ & $59.63$ & $59.63$ & $[59.476, 59.872]$  \\
$50$ & $69.22$ & $69.23$ & $69.70$ & $69.56$ & $69.57$ & $69.51$ & $69.57$ & $ [69.560, 69.945]$  \\
$100$ & $83.14$ & $83.18$ & $83.29$ & $83.33$ & $83.37$ & $83.39$ & $83.16$ & $ [83.357, 83.862]$  \\
$200$ & $97.21$ & $97.07$ & $97.31$ & $97.43$ & $97.41$ & $97.46$ & $97.21$ & $[97.381, 97.889]$  \\
$500$ & $116.13$ & $116.07$ & $116.17$ & $116.31$ & $116.31$ & $116.36$ & $116.14$ & $[116.210, 116.685]$  \\
$750$ & $124.56$ & $124.56$ & $124.61$ & $124.72$ & $124.73$ & $124.78$ & $124.59$ \\
$1000$ & $130.65$ & $130.63$ & $130.66$ & $130.78$ & $130.83$ & $130.84$ & $130.67$ \\
\end{tabular}
\caption{$n = 9$, $T=3$, $r=0.05$, $\delta = 0.1$, $\sigma=0.2$, $\rho=0$, $S_0^j=100$, $K=100$, $\omega_j=1$, $m=10^6$, $m_{\mathrm{resim}}=10^6$ not using reordering \\ From: \cite{andersen2004primal, becker2019deep}}
\label{tbl:experiment_max_call}
\end{table}
\begin{table}[!ht]
\centering
\begin{tabular}{ c | c c c c c c c | c}
$d$ & \multicolumn{7}{c}{p} & $V_\textrm{ref}$ \\
 \hline
 & $1$ & $2$ & $3$ & $4$ & $5$ & $6$ & $7$ & \\
\hline
$2$ & $13.67$ & $13.76$ & $13.82$ & $11.63$ & $13.84$ & $13.84$ & $13.85$  & $13.902$ \\
$3$ & $18.39$ & $18.51$ & $18.60$ & $18.61$ & $18.61$ & $18.62$ & $18.62$ & $18.69$ \\
$5$ & $25.83$ & $26.01$ & $26.06$ & $26.07$ & $26.07$ & $26.07$  & $26.07$ & $[26.115, 26.164]$\\
$10$ & $38.08$ & $38.24$ & $38.29$ & $38.31$ & $38.31$ & $38.30$ & $38.30$ & $[38.300, 38.367]$ \\
$20$ & $51.48$ & $51.66$ & $51.71$ & $51.71$ & $51.71$ & $51.71$ & $51.71$ & $[51.549, 51.803]$ \\
$30$ & $59.50$ & $59.68$ & $59.71$ & $59.71$ & $59.72$ & $59.72$ & $59.72$ & $[59.476, 59.872]$  \\
$50$ & $69.58$ & $69.78$ & $69.80$ & $69.81$ & $69.81$ & $69.81$ & $69.81$ & $ [69.560, 69.945]$  \\
$100$ & $83.45$ & $83.65$ & $83.67$ & $83.67$ & $83.67$ & $83.66$ & $83.66$ & $ [83.357, 83.862]$  \\
$200$ & $97.56$ & $97.69$ & $97.70$ & $97.70$ & $97.70$ & $97.69$ & $97.69$ & $[97.381, 97.889]$  \\
$500$ & $116.45$ & $116.56$ & $116.56$ & $116.56$ & $116.56$ & $116.50$ & $116.52$ & $[116.210, 116.685]$  \\
$750$ & $124.91$ & $124.98$ & $124.99$ & $124.98$ & nan & nan & nan  \\
$1000$ & $130.96$ & $131.06$ & $131.05$ & nan & nan & nan & nan \\
\end{tabular}
\caption{$n = 9$, $T=3$, $r=0.05$, $\delta = 0.1$, $\sigma=0.2$, $\rho=0$, $S_0^j=100$, $K=100$, $\omega_j=1$, $m=10^6$, $m_{\mathrm{resim}}=10^6$ using reordering \\ From: \cite{andersen2004primal, becker2019deep} }
\label{tbl:experiment_max_call_sort}
\end{table}

\FloatBarrier

\section*{Acknowledgements}
Christian Bayer gratefully acknowledges support by the DFG cluster of excellence MATH+, project AA4-2.
Leon Sallandt acknowledges support from the Research Training Group ``Differential Equation- and Data-driven Models in Life Sciences and Fluid Dynamics: An Interdisciplinary Research Training Group (DAEDALUS)'' (GRK 2433) funded by the German Research Foundation (DFG).
Philipp Trunschke acknowledges support by the Berlin International Graduate School in Model and Simulation based Research (BIMoS).
We thank Max Pfeffer and Reinhold Schneider for fruitful discussions.

\printbibliography

\end{document}